\documentstyle[preprint,aps]{revtex}
\begin{document}
\draft
\preprint{  }
\title{Time-dependent transport in interacting
and non-interacting mesoscopic systems}
\author{Antti-Pekka Jauho}
\address{
MIC, Technical University of Denmark, DK-2800 Lyngby, Denmark, and\\
Nordita, Blegdamsvej 17, DK-2100 Copenhagen {\O}, Denmark
}

\author{Ned S. Wingreen}
\address{NEC Research Institute, 4 Independence Way, Princeton, NJ
08540}\date{\today}

\author{Yigal Meir}
\address{Department of Physics, University of California at Santa
Barbara, CA 93106}

\maketitle

\begin{abstract}
We consider a mesoscopic region coupled to two
leads under the influence of external
time-dependent voltages. The time dependence is coupled to source
and drain contacts, the gates controlling the tunnel-barrier heights, or
to the gates which define the mesoscopic region.
We derive, with the Keldysh nonequilibrium Green function
technique, a formal expression for the fully nonlinear, time-dependent
current through the system.  The analysis admits arbitrary
interactions in the mesoscopic region, but the leads are
treated as noninteracting.
For proportionate coupling to the leads, the
time-averaged current is simply the integral
between the chemical potentials of the time-averaged
density of states, weighted by the coupling to the leads, in close analogy
to the time-independent result of Meir and Wingreen
(Phys. Rev. Lett. {\bf 68}, 2512 (1992)).  Analytical and numerical
results for the exactly solvable non-interacting resonant-tunneling system
are presented.
Due to the coherence between the leads and the resonant site,
the current
does not follow the driving signal adiabatically:
a "ringing" current is found as a response to a voltage pulse, and
a complex time-dependence results in the case of harmonic driving
voltages.
We also establish a connection to recent linear-response calculations, and
to earlier studies of electron-phonon scattering effects in resonant
tunneling.
\end{abstract}

\section{Introduction}
The hallmark of mesoscopic phenomena is the phase coherence of the
charge carriers, which is maintained over a significant part of
the transport process.  The interference effects resulting from this
phase coherence are reflected in a number of experimentally measurable
properties.
For example, phase coherence is central
to the Aharonov-Bohm effect,\cite{alw} Universal conductance
fluctuations,\cite{alw} and weak localization,\cite{lr} and
can be affected by external controls such as temperature
or magnetic field.
The study of stationary mesoscopic physics is now a
mature field, and in this work we focus on an alternative way of affecting the
phase coherence: external {\it time-dependent} perturbations.
The interplay of external time dependence and phase coherence
can be phenomenologically understood as follows.  If the single-particle
energies acquire a time dependence, then the wave functions have
an extra phase factor, $\psi \sim \exp(-i\int^t dt'\epsilon(t'))$.
For a uniform system such an overall phase factor is of no
consequence. However, if the external time dependence is
different in different parts of the system, and the particles
can move between these regions (without being 'dephased'
by inelastic collisions), the phase difference becomes
important.

The interest in time-dependent mesoscopic phenomena stems from
recent progress in several experimental
techniques.\cite{kirkreed}  Time dependence
is a central ingredient in many different experiments,
of which we mention the following:
(i) {\it Single-electron pumps and turnstiles.}
Here time-modified gate signals move electrons one by one through
a quantum dot, leading to a current which is proportional to the
frequency of the external signal.  These structures have considerable
importance as current standards.  The Coulombic repulsion
of the carriers in the central region
is crucial to the operational principle of these devices, and
underlines the fact that extra care must be paid to interactions when
considering time-dependent transport in mesoscopic systems.
(ii) {\it ac response and transients in resonant-tunneling devices.}
Resonant tunneling devices (RTD) have a number of applications as
high-frequency amplifiers or detectors.  For the device engineer
a natural approach would be to model these circuit elements with
resistors, capacitances, and inductors.  The question then arises
as to what, if any, are the appropriate 'quantum' capacitances and inductances
one should ascribe to these devices.  Answering this question requires
the use of time-dependent quantum-transport theory.
(iii) {\it Interaction with laser fields.}  Ultrashort laser pulses
allow the study of short-time dynamics of charge carriers.  Here again,
coherence and time dependence combine with the necessity of treating
interactions.

A rigorous discussion of transport in an interacting mesoscopic system
requires a formalism which is capable of including explicitly the
interactions.  Obvious candidates for such a theoretical tool
are various techniques based on Green functions.  Since many
problems of interest involve systems far from equilibrium,
we cannot use linear-response methods, such as those based
on the Kubo formula, but must use an approach capable of
addressing the full nonequilibrium situation.  The nonequilibrum
Green function techniques, as developed about thirty years
ago by Kadanoff and Baym,\cite{KB} and by Keldysh,\cite{Keldysh}
have during the recent years gained increasing attention in
the analysis of transport phenomena in mesoscopic semiconductors
systems.\cite{Caroli}  In particular, the {\it steady state} situation.
has been addressed by a large number of papers
\cite{MW92,Davies,Anda,HDW,MWL,Lake}  Among the central results obtained in
these papers
is that that under certain conditions (to be discussed below)
a Landauer type conductance formula\cite{Landauer} can be
derived.  This is quite appealing in view of the wide spread
success of conductance formulas  in the analysis of transport in
mesoscopic systems.

Considerably fewer studies have been reported where an
explicit time dependence is an essential feature.
We are aware of an early paper in surface physics,\cite{Blandin}
but only in the recent past have  groups
working in mesoscopic physics addressed
this problem.\cite{chenting,LangNord,WJM93,Bruder,Runge,Pastawski}
The work reported in this paper continues
along these lines: we give the full details and expand
on our short communication.\cite{WJM93}

Our main formal
result from the nonequilibrium Green function approach
is  a general expression for the time-dependent
current flowing from non-interacting leads to an interacting
region.  As we will discuss in Section II,
the time dependence enters through the
self-consistent parameters defining the model.  We show that under certain
restrictions, to be specified below, a Landauer-like formula
can be obtained for the {\it time-averaged} current.  To illustrate
the utility of our approach we give results for an exactly
solvable non-interacting case, which displays an interesting,
and experimentally measurable, nonadiabatic behavior.  We also
establish a link between the present formulation and recently published results
for linear-response and electron-phonon interactions,
obtained by other techniques.

The paper is organized as follows.
We examine in Section II the range of experimental parameters in
which we expect our theoretical formulation to be valid.
In Section III we
briefly review the physics behind the
nonequilibrium Green function technique of Keldysh, and Baym and Kadanoff,
which is our main theoretical tool, and then introduce the specific
model Hamiltonians used in this work.  We derive the central formal results for
the time-dependent
current in Section IV. We also derive, under special restrictions,
a Landauer-like formula for the average current. In Section V
we apply the general formulae
to an explicitly solvable resonant-tunneling model.  Both analytical and
numerical results are
presented.  We also show that the linear ac-response
results of Fu and Dudley\cite{fududley} are contained
as a special case of the exact results of this section.  In Section VI
we illustrate the utility of our formulation by
presenting a much simplified derivation of
Wingreen et al's\cite{ned} results on resonant
tunneling in the presence of electron-phonon interactions.
Appendix A summarizes some of the central
technical properties of the Keldysh technique:
we state the definitions, give the basic
equations,  and provide the analytic continuation
rules employed below.
In Appendices B and C we present  proofs
for certain statements made in the main text, and, finally, in
Appendix D we describe some
transformations which facilitate
numerical evaluation of the time-dependent
current.

\section{Applicability to experiments}

A central question one must address is:
under which conditions
are the non-equilibrium techniques, applied successfully to the
steady-state problem, transferrable to time-dependent situations,
such as the experiments mentioned above?

The time-dependent problem has to be formulated carefully,
particularly with respect to the leads. It is essential
to a Landauer type of approach, that the electrons in the
leads be non-interacting.
In practice, however, the electrons in the leads near the
mesoscopic region contribute to the self-consistent potential.
We approach this problem
by dividing the transport physics in two steps:\cite{Buettiker} (i) the
self-consistent determination of charge pile-up and depletion
in the contacts,
the resulting barrier heights, and single-particle energies in
the interacting region, and (ii) transport in a system
defined by these self-consistent parameters.
Step (i) requires a capacitance calculation
for each specific geometry,\cite{Buettiker} and we do
not address it in this paper.
Instead, we
assume the results of (i) as time-dependent input parameters and give
a full treatment of the transport
through the mesoscopic region (ii). In practice, the interactions
in the leads are absorbed into a time-dependent potential and from
then on the electrons in the leads are treated as non-interacting.
This means that when relating our results to actual experiments
some care must be exercised. Specifically, we calculate only the
current flowing into the mesoscopic region, while the total time-dependent
current measured in the contacts includes contributions from
charge flowing in and out of accumulation and depletion regions
in the leads. In the {\it time-averaged} (dc) current, however, these
capacitive contributions vanish and the corresponding time-averaged
theoretical formulae, such as Eq.(\ref{jave}), are directly relevant to
experiment.
It should be noted, though, that these capacitive currents may
influence the effective time-dependent parameters in step (i) above.

Let us next estimate the frequency limits that restrict
the validity of
our approach.
Two criteria must be satisfied.  First,
the driving frequency must be sufficiently slow that the
applied bias is dropped entirely across the tunneling structure.
When a bias is applied to a sample, the electric
field in the leads can only be screened if the driving
frequency is smaller than the plasma frequency, which is tens of
THz in typical doped semiconductor samples.
For signals slower than this, the bias is
established entirely across the tunneling structure by accumulation and
depletion of charge near the barriers.
The unscreened Coulomb interaction
between net excess charge is quite strong, and hence
the bias across a tunneling structure is caused by a relatively
small excess of charge in accumulation and depletion layers.
The formation of these layers then causes a rigid shift
(see Eq.(\ref{glead}) below) of the bottom
of the conduction band deeper in the leads, which is the origin
of the rigid shift of energy levels in our treatment of
a time-dependent bias.

The second frequency limit on our approach
is that the build-up of electrons required for the
formation of the accumulation and depletion layers must not
significantly disrupt the coherent transport of electrons
incident from the leads.
One way to quantify this is to ask --  what is the probability
that an electron incident from the leads participates in the
build-up of charge associated with a time-dependent bias?
This probability will be the ratio of the net current
density flowing into the accumulation
region to the total incident flux of electrons. For a
three dimensional double-barrier
resonant-tunneling structure (see Fig.\ \ref{fig1})
the ac-current charging the accumulation
layer is
$I_{\mathrm{acc}}^{\mathrm{rms}} = 2 \pi \nu C
V^{\mathrm{rms}}/A, $
where $\nu$ is the driving frequency, $C$ is the capacitance,
$V^{\mathrm{rms}}$ is the applied bias, and $A$ is the area.
In comparison, the total incident flux is
$ I_{\mathrm{inc}} = 3/8\, e n v_{\mathrm{F}}.$
Using the parameters appropriate for a typical
experiment (we use that of Brown {\it et al.}\cite{Brown}),
we find that  up to 10 THz the
probability of an electron participating in the charge build-up
is only 1\%.
Summarizing, these estimates indicate that our approach should be accurate
up to frequencies of tens of THz, which are large
by present experimental standards [Ref. ?], and consequently
the analysis presented in what follows should be valid for most
experimental situations.

\section{Theoretical tools, and the model}

\subsection{Baym-Kadanoff-Keldysh nonequilibrium
techniques}

Here we wish to outline the physical background behind the
Keldysh formulation, and in particular its connection to
tunneling physics.
Readers interested in technical details should consult
any of the many available review articles, such as
Refs.[\onlinecite{Langreth,Rammer,Jauho}].
The basic difference between
construction of
equilibrium
and nonequilibrium perturbation schemes is that in nonequilibrium
one cannot assume that the system returns to its ground
state (or a thermodynamic equilibrium state at finite
temperatures) as $t\to +\infty$.  Irreversible
effects break the symmetry between $t=-\infty$ and
$t=+\infty$, and this symmetry is heavily
exploited in the derivation of the equilibrium perturbation
expansion.  In nonequilibrium situations one can circumvent
this problem by allowing the system to evolve from
$-\infty$ to the moment of interest (for definiteness,
let us call this instant $t_0$), and then continues the
time evolvement from $t=t_0$ back to
$t=-\infty$.\cite{Schwinger}
(When dealing with quantities which depend
on two time variables, such as Green functions, the time
evolution must be continued to the later time.)
The advantage of this
procedure is that all expectation values are defined
with respect to a well defined state, i.e. the state
in which the system was prepared in the remote past.  The
price  is that one must treat the
two time branches on an equal footing (See Fig.\ \ref{fig2}).

A typical object of interest would be a two time Green function
(see Appendix A);
the two times can be located on either of the two branches
of the complex time path (e.g., $\tau$ and $\tau'$ in
Fig.\ \ref{fig2}).
One is thus led to consider
$2\times2$ Green function matrices, and the various
terms in the perturbation theory can be evaluated by
matrix multiplication.  Since the internal time-integrations
run over the complex time path,
a method of book-keeping for the time-labels is required,
and there are various ways of doing this.  In the
present work we employ a version of the
Keldysh technique.

In the context of tunneling problems the time-independent Keldysh formalism
works as follows.  In the remote past the contacts (i.e. the
left and right lead) and the central region are decoupled,
and each region is in thermal equilibrium.
The equilibrium distribution functions for the three regions
are characterized by their respective chemical potentials;
these do not have to coincide nor are the differences between
the chemical potentials necessarily small.  The couplings between the
different regions
are then established and treated as perturbations
via  the standard techniques of perturbation
theory, albeit on the two-branch time contour.
It is important
to notice that the couplings do not have to be small,
{\it e.g.} with respect level spacings or $k_{\mathrm{B}} T$, and typically
must be treated to all orders.

The time-dependent case can be treated similarly.
Before the the couplings between the various regions are turned on,
the single-particle energies acquire
rigid time-dependent shifts,
which, in the case of the non-interacting contacts, translate
into extra phase factors for the propagators (but not in
changes in occupations).
The perturbation theory with respect to the couplings has
the same diagrammatic structure as in the stationary case.
The calculations, of course, become more complicated because
of the broken time translational invariance.

\subsection{Model Hamiltonian}
We split the total Hamiltonian in three pieces:
$H=H_c+H_T+H_{\mathrm{cen}}$, where $H_c$ describes the contacts,
$H_T$ is the tunneling coupling between contacts and the
interacting region, and $H_{\mathrm{cen}}$ models the interacting
central region, respectively.  Below we discuss each of these terms.
\subsubsection{Contacts, $H_c$}
Guided by the typical experimental geometry in which
the leads rapidly broaden into metallic contacts, we view
electrons in the leads as non-interacting except for an
overall self-consistent potential.
Physically, applying a time-dependent bias (electrostatic-potential
difference) between the source and drain contacts means that the
single-particle energies become time-dependent:
$\epsilon_{k\alpha}^0 \to
\epsilon_{k\alpha}(t)=\epsilon_{k\alpha}^0+\Delta_{\alpha}(t)$ (here $\alpha$
labels the channel in the left ($L$) or right ($R$) lead).
The occupation of
each state $k\alpha$, however, remains unchanged.
The occupation, for each contact, is determined by an
equilibrium distribution function established in
the distant past, before the time dependence or tunneling
matrix elements are turned on.
Thus, the contact Hamiltonian is
\begin{equation}\label{hlead}
H_c =  \sum_{k,\alpha\in L,R} \epsilon_{k\alpha}(t)
{\mathbf{c}}_{k\alpha}^{\dagger} {\mathbf{c}}_{k\alpha}\;,
\end{equation}
and the exact time-dependent Green functions in the leads for
the uncoupled system are
\begin{eqnarray}\label{glead}
g_{k\alpha}^<(t,t') && \equiv
i\langle{\mathbf{c}}_{k\alpha}^{\dagger}(t'){\mathbf{c}}_{k\alpha}(t)\rangle
\nonumber\\ && =
i f(\epsilon^0_{k\alpha})
\exp\big [ -i \int_{t'}^t dt_1\epsilon_{k\alpha}(t_1)\big ] \nonumber\\
g_{k\alpha}^{r,a}(t,t') && \equiv \mp i\theta(\pm t \mp t')
\langle\lbrace{\mathbf{c}}_{k\alpha}(t),
{\mathbf{c}}_{k\alpha}^{\dagger}(t')\rbrace\rangle
\nonumber\\ && =
\mp i\theta(\pm t \mp t')
\exp\big [ -i \int_{t'}^t dt_1\epsilon_{k\alpha}(t_1)\big ] \;.
\end{eqnarray}
One should note that our model for $g^<$ differs from the choice
made in the recent study of Chen and Ting:\cite{chenting}
these authors allow the electrochemical potential in the distribution
function $f$  to vary with time:
$\mu_L-\mu_R= e[V+U(t)]$, where $U(t)$ is the ac signal.
This assumption implies that the {\it total number} of electrons in the
contacts varies with time.  This behavior is inconsistent of what happens
in real devices: it is
only the relatively small number of electrons in the accumulation/depletion
layers that is time-dependent.  In addition to the unphysical charge
pile-up in the contacts,
the model of Chen
and Ting leads to an instantaneous loss of phase-coherence in
the contacts, and hence does not display any of the
interesting interference phenomena predicted by our phase-conserving
model.

\subsubsection{Coupling between leads and central region, $H_T$}
The coupling between the leads and the central (interacting) region
can be modified with time-dependent gate voltages, as is the
case in single-electron pumps.  The precise functional form
of the time-dependence is determined by
the detailed geometry and by the self-consistent response of
charge in the contacts to external driving.  We assume that
these parameters are known, and simply write
\begin{equation}\label{ht}
H_T=
\sum_{k,\alpha\in L,R\atop n}
[V_{k\alpha,n}(t){\mathbf{c}}_{k\alpha}^{\dagger} {\mathbf{d}}_n +
{\mathrm{h.c.}} ]
\;.
\end{equation}
Here $\{{\mathbf{d}}^{\dagger}_n\}$ and $\{{\mathbf{d}}_n\}$
form a complete orthonormal
set of single-electron creation and annihilation operators in
the interacting region.

\subsubsection{The central region Hamiltonian $H_{\mathrm{cen}}$}
The form chosen for $H_{{\mathrm cen}}$ in the
central interacting region depends on geometry and on the physical
behavior being investigated.  Our results
relating the current to local properties,
such as densities of states and Green functions,
are valid generally.  To make
the results more concrete, we will discuss two
particular examples in detail.
In the first, the central region is taken to
consist of noninteracting, but time-dependent levels,
\begin{equation}\label{nonint}
H_{\mathrm{ cen}}=\sum_m\epsilon_m(t)
{\mathbf{d}}_m^{\dagger}{\mathbf{d}}_m
\end{equation}
Here ${\mathbf{d}}_m^{\dagger}({\mathbf{d}}_m)$ creates (destroys) an electron
in state $m$.  The choice (\ref{nonint}) represents
a simple model for time-dependent resonant tunneling.
Below we shall present general results for an arbitrary number
of levels, and analyze the case of a single level
in detail. The latter is
interesting both as an exactly solvable example, and for
predictions of coherence effects in time-dependent experiments.

The second example we will discuss is resonant tunneling
with electron-phonon interaction,
\begin{equation}\label{helph}
H_{\mathrm{ cen}}^{\mathrm{ el-ph}} =
\epsilon_0 {\mathbf{d}}^{\dagger}{\mathbf{d}} + {\mathbf{d}}^{\dagger}
{\mathbf{d}}\sum_{{\mathbf q}}
M_{{\mathbf q}}[{\mathbf{a}}^{\dagger}_{{{\mathbf q}}}+{\mathbf{a}}_{-{\mathbf
q}}]\;,
\end{equation}
In the above, the first term
represents a single site, while the second term represents
interaction of an electron on the site with
phonons: ${\mathbf{a}}^{\dagger}_{{\mathbf  q}}({\mathbf{a}}_{{\mathbf q}})$
creates (destroys) a phonon in mode ${\mathbf  q}$,
and $M_{{\mathbf q}}$ is the interaction matrix element.
The full Hamiltonian of the system must also include
the free-phonon contribution
$H_{{\mathrm ph}}=\sum_{{\mathbf q}}\hbar
\omega_{{\mathbf q}} {\mathbf{a}}^{\dagger}_{{\mathbf q}}
{\mathbf{a}}_{{\mathbf q}}$.
This example, while not exactly solvable, is helpful
to show how interactions influence the
current.  Furthermore, we can directly compare to
previous time-independent results \cite{ned} using
(\ref{helph}) to demonstrate the power of the present formalism.

\section{Time-dependent current and Keldysh Green functions}

\subsection{General expression for the current}

The current from the left contact
through left  barrier to the central region can be calculated from
the time evolution of the occupation number operator of the
left contact:
\begin{equation}
J_L(t)= -e \langle\dot N_L\rangle =
-{ie\over\hbar}\langle [H,N_L]\rangle\;,
\end{equation}
where $N_L = \sum_{k,\alpha\in L}
{{\mathbf c}}_{k\alpha}^{\dagger} {\mathbf{c}}_{k\alpha}$
and $H=H_c+H_T+H_{\mathrm{ cen}}$.
Since $H_c$ and $H_{\mathrm{ cen}}$ commute with $N_L$, one readily finds
\begin{equation}
J_L = {ie\over\hbar} \sum_{k,\alpha\in L\atop n}
[V_{k\alpha,n}\langle{{\mathbf c}}^{\dagger}_{k\alpha}{{\mathbf d}}_n\rangle
-V^*_{k\alpha,n}\langle{{\mathbf d}}^{\dagger}_n{{\mathbf
c}}_{k\alpha}\rangle]\;.
\end{equation}
Now define two Green functions:
\begin{eqnarray}\label{gnkalpha}
G_{n,k\alpha}^<(t,t') && \equiv i\langle {{\mathbf c}}^{\dagger}_{k\alpha}(t')
{{\mathbf d}}_n(t)\rangle \\
G_{k\alpha,n}^<(t,t') && \equiv i\langle{{\mathbf d}}_n^{\dagger}(t')
{{\mathbf c}}_{k\alpha}(t)\rangle \;.
\end{eqnarray}
Using $G_{k\alpha,n}^<(t,t)
= -\left [ G_{n,k\alpha}^<(t,t) \right ]^*$, and inserting the time
labels,  the current can be expressed as
\begin{equation}\label{current}
J_L(t)= {2e\over\hbar} {\mathrm{ Re}} \Bigl\lbrace \sum_{k,\alpha\in L\atop n}
V_{k\alpha,n}(t)
G_{n,k\alpha}^<(t,t)\Bigr\rbrace \;.
\end{equation}
One next needs an expression for $G_{n,k\alpha}^<(t,t')$.  For the present
case, with non-interacting leads, a general relation
for the contour-ordered Green function
$G_{n,k\alpha}(\tau,\tau')$ can be derived
rather easily (either with the equation-of-motion technique, or by
a direct expansion of the $S$-matrix; the details are given in Appendix B),
and the result is
\begin{equation}\label{Gnktau}
G_{n,k\alpha}(\tau,\tau') =  \sum_m \int d\tau_1 G_{nm}(\tau,\tau_1)
V^*_{k\alpha,m}(\tau_1)g_{k\alpha}(\tau_1,\tau')\;.
\end{equation}
Here $G_{nm}(\tau,\tau_1)$ is the contour-ordered Green function for the
central
region, and
the $\tau$-variables are now defined on the contour of Fig.\ \ref{fig2}.
Note that the time-dependence of the tunneling matrix elements and
single-particle
energies has broken the time-translational invariance.
The analytic continuation rules (\ref{rules}) of Appendix A can
now be applied, and we find
\begin{eqnarray}\label{dyseq}
G^<_{n,k\alpha}(t,t') = && \sum_m \int dt_1V^*_{k\alpha,m}(t_1)
[ G^r_{nm}(t,t_1)g^<_{k\alpha}(t_1,t')\nonumber\\
+ && G^<_{nm}(t,t_1)g^a_{k\alpha}(t_1,t')] \;,
\end{eqnarray}
where the Green functions $g^{<,a}$ for the leads are defined in (\ref{glead})
above.
Combining (\ref{glead}),(\ref{current}), and (\ref{dyseq}), yields
\begin{eqnarray}\label{cur1}
&& J_L(t) = -{2e\over\hbar} {\mathrm{ Im}}\Bigl\lbrace
\sum_{k,\alpha\in L\atop {n,m}} V_{k\alpha,n}(t)
\int^t_{-\infty} dt_1 e^{i\int_{t_1}^tdt_2\epsilon_{k\alpha}(t_2)}
\nonumber\\
&&\times  V^*_{k\alpha,m}(t_1) [G^r_{nm}(t,t_1)f_L(\epsilon_{k\alpha}) +
G^<_{nm}(t,t_1)]\Bigr\rbrace  \;.
\end{eqnarray}
The discrete sum over $k$ in $\sum_{k\alpha}$
can be expressed in terms of densities of states in the leads:
$\int d\epsilon\rho_{\alpha}(\epsilon)$.
Then it is useful to define
\begin{eqnarray}\label{gamma}
\big [ \Gamma^L(\epsilon,t_1,t) \big ]_{mn}= &&
2\pi \sum_{\alpha\in L}  \rho_{\alpha}(\epsilon)
V_{\alpha,n}(\epsilon,t)V^*_{\alpha,m}(\epsilon,t_1) \nonumber\\
&& \times \exp[i\int_{t_1}^t dt_2\Delta_{\alpha}(\epsilon,t_2)]\;,
\end{eqnarray}
where $V_{k\alpha,n}=V_{\alpha,n}(\epsilon_k)$.
In terms of this generalized line-width function (\ref{gamma}), the general
expression
for the current is
\begin{eqnarray}\label{curfin}
J_L(t)= && -{2e\over\hbar}
\int_{-\infty}^t dt_1 \int {d\epsilon\over2\pi}
{\mathrm{ ImTr}} \Bigl\lbrace
{\mathrm{ e}}^{-i\epsilon(t_1-t)}
{{\mathbf \Gamma}}^L(\epsilon,t_1,t) \nonumber\\
&& \times [{{\mathbf G}}^<(t,t_1)
+f_L(\epsilon){{\mathbf G}}^r(t,t_1)]\Bigr\rbrace\;.
\end{eqnarray}
Here the bold-face notation indicates that the level-width
function $\mathbf\Gamma$ and the central-region Green functions
${\mathbf{G}}^{<,r}$ are matrices in the central-region
indeces $m,n$.
An analogous formula applies for $J_R(t)$, the current flowing
into the central region through the right barrier.

This is the central formal result of this work, and the remainder
of this paper is devoted to the analysis and evaluation of Eq.(\ref{curfin}).
The current is expressed in terms of local quantities: Green functions
of the central region.
The first term in Eq.(\ref{curfin}), which is proportional to the
lesser function $G^<$, suggests an interpretation as the
out-tunneling rate (recalling ${\mathrm{Im}}\,G^<(t,t)=N(t)$).  Likewise, the
second term, which is proportional to the occupation in the
leads and to the density of states in the
central region, can be associated to the in-tunneling rate.
However, one should bear in mind that
all Green functions in  Eq.(\ref{curfin})
are to be calculated in the presence of tunneling.
Thus, $G^<$ will depend
on the occupation in the leads. Furthermore, in
the presence of interactions $G^r$ will depend on the central
region occupation.
Consequently, the current can be a non-linear
function of the occupation factors.
This issue has recently been discussed also
by other authors.\cite{Lakeetal}

\subsection{Time-independent case}

\subsubsection{General expression}

In the time-independent limit the line-width function simplifies:
${\mathbf \Gamma}(\epsilon,t_1,t)\to{\mathbf \Gamma}(\epsilon)$,
and the $t_1$-integrals in Eq.(\ref{curfin}) can be performed:
\begin{eqnarray}
\int_{-\infty}^t dt_1 && \int {d\epsilon\over2\pi}
{\mathrm{ImTr}} \Bigl\lbrace
{\mathrm{ e}}^{-i\epsilon(t_1-t)}
{\mathbf \Gamma}^L(\epsilon) {\mathbf G}^<(t-t_1)\Bigr\rbrace \nonumber\\
&& =
- {i\over 2} \int {d\epsilon\over2\pi}
{\mathrm{ Tr}}\lbrace{\mathbf \Gamma}^L(\epsilon){\mathbf G}^<(\epsilon)
\rbrace\;,
\end{eqnarray}
and
\begin{eqnarray}
\int_{-\infty}^t && dt_1  \int {d\epsilon\over2\pi}
{\mathrm{ImTr}} \Bigl\lbrace
{\mathrm {e}}^{-i\epsilon(t_1-t)}
{\mathbf \Gamma}^L(\epsilon)
f_L(\epsilon){\mathbf G}^r(t-t_1)\Bigr\rbrace \nonumber\\
&& =
- {i\over 2} \int {d\epsilon\over2\pi}
{\mathrm{ Tr}}\Bigl\lbrace{\mathbf \Gamma}^L(\epsilon)f_L(\epsilon)
\bigl[{\mathbf G}^r(\epsilon)-{\mathbf G}^a(\epsilon)\bigr]\Bigr\rbrace\;.
\end{eqnarray}
When these expressions are substituted to Eq.(\ref{curfin}),
the
current from left (right) contact to central region becomes
\begin{eqnarray}\label{leftdccur}
J_{L(R)}= && {ie\over\hbar}\int{d\epsilon\over 2\pi}
{\mathrm{ Tr}}\Bigl\lbrace{\mathbf \Gamma}^{L(R)}(\epsilon)
\bigl ( {\mathbf G}^<(\epsilon)\nonumber\\
&& +
f_{L(R)}(\epsilon)[{\mathbf G}^r(\epsilon)-{\mathbf G}^a(\epsilon)]\bigr )
\Bigr\rbrace\;.
\end{eqnarray}
In steady state, the current will be uniform, so that
$J = J_L=-J_R$ , and one can symmetrize the current:
$J=(J_L+J_L)/2=(J_L-J_R)/2$.  Using Eq.(\ref{leftdccur}) leads to
the general expression for the d.c. current:
\begin{eqnarray}\label{Ned92}
J= && {ie\over2\hbar}
\int {d\epsilon\over 2\pi}
{\mathrm{ Tr}} \bigl \lbrace
[{\mathbf \Gamma}^L(\epsilon) - {\mathbf \Gamma}^R(\epsilon)]
{\mathbf G}^<(\epsilon)\nonumber\\
 + && [f_L(\epsilon){\mathbf \Gamma}^L(\epsilon)
-f_R(\epsilon){\mathbf \Gamma}^R(\epsilon)]
[{\mathbf G}^r(\epsilon)-{\mathbf G}^a(\epsilon)]\bigr\rbrace\;.
\end{eqnarray}
This result was reported in Ref.\onlinecite{MW92}, and applied to the
out-of-equilibrium Anderson impurity problem.

\subsubsection{Proportionate coupling}
If the left and right line-width functions are proportional to each other,
i.e. ${\mathbf \Gamma}^L(\epsilon) =
\lambda {\mathbf \Gamma}^R(\epsilon)$, further simplification can
be achieved.  We observe that the
current can be written as $J\equiv xJ_L-(1-x)J_R$, which gives,
using Eq.(\ref{leftdccur}),
\begin{eqnarray}
J= && {ie\over\hbar}\int {d\epsilon\over 2\pi}{\mathrm{ Tr}}\Bigl\{
{\mathbf \Gamma}^R(\epsilon)\bigl [ (\lambda x-(1-x)){\mathbf G}^<(\epsilon)
 \nonumber\\
&& +
(\lambda xf_L-(1-x) f_R)({\mathbf G}^r(\epsilon)-{\mathbf G}^a)(\epsilon)
\bigr ] \Big\} \;.
\end{eqnarray}
The arbitrary parameter $x$
is now fixed
so that the first term vanishes, i.e $x=1/(1+\lambda)$, which
results in
\begin{eqnarray}\label{jdcprop}
J= && {ie\over\hbar}\int{d\epsilon\over 2\pi}[f_L(\epsilon)-f_R(\epsilon)]
\nonumber\\
\times &&{\mathrm{ Tr}}
\Bigl\lbrace {{\mathbf \Gamma}^L(\epsilon){\mathbf \Gamma}^R(\epsilon)\over
{\mathbf \Gamma}^L(\epsilon)+{\mathbf \Gamma}^R(\epsilon)}
\bigl({\mathbf G}^r(\epsilon)-{\mathbf G}^a(\epsilon)\bigr )\Bigr\rbrace
\;.
\end{eqnarray}
The ratio is well-defined because
the ${\mathbf{\Gamma}}$-matrices  are proportional.
The difference between the retarded and advanced Green functions
is essentially the density of states. Despite of the apparent
similarity
of (\ref{jdcprop}) to the Landauer formula, it is important to bear in
mind that in general there is no immediate connection between
the quantity
${\mathrm{ Tr}}
\Bigl\lbrace \bigl( {\mathbf \Gamma}^L(\epsilon){\mathbf \Gamma}^R(\epsilon) /
\bigl [ {\mathbf \Gamma}^L(\epsilon)+{\mathbf \Gamma}^R(\epsilon)\bigr ]\bigr )
\bigl [ {\mathbf G}^r(\epsilon)-{\mathbf G}^a(\epsilon)\bigr ]\Bigr\rbrace
$, and
the transmission coefficient.
In particular, when inelastic scattering
is present, we do not believe that such a connection exists.
In Section V, where we analyze a non-interacting central
region, a connection with the transmission coefficient {\it can}
be established.
Further, in the next section we shall see how
an analogous result can be derived for the average of the
time-dependent current.

\subsection{Average current}\label{avecur}
In analogy with the previous subsection, where we found a compact
expression for the current for the case of proportionate coupling,
the time-dependent case allows further simplification, if
assumptions are made on the line-width functions.  In this
case, we assume a
generalized proportionality condition:
\begin{equation}\label{propcond}
{\mathbf \Gamma}^L(\epsilon,t_1,t)
= \lambda {\mathbf \Gamma}^R(\epsilon,t_1,t)\;.
\end{equation}
One should note that in general this condition can be satisfied
only if
$\Delta^L_{\alpha}(t)=\Delta^R_{\alpha}(t)=\Delta(t)$.
However, in the Wide-Band Limit (WBL), to be considered in detail below,
the time-variations of the energies
in the leads do not have to be equal.

We next consider the occupation of
the central region
$N(t) = \sum_m\langle {\mathbf d}_m^{\dagger}(t){\mathbf d}_m(t)\rangle$
and apply the continuity equation:
\begin{equation}\label{conteq}
e{dN(t) \over dt} = J_R(t)+J_L(t)\;,
\end{equation}
which allows one to write for arbitrary x:
\begin{equation}
J_L(t)=xJ_L(t)+(1-x)[e{dN(t) \over dt}-J_R(t)]\;.
\end{equation}
Choosing now $x\equiv 1/(1+\lambda)$ leads to
\begin{eqnarray}\label{eqn}
&&J_L(t)  = \Big ( {\lambda\over 1+\lambda}  \Big )
\biggl [ e{dN\over dt}
- {2e\over\hbar}{\mathrm{ImTr}} \Bigl\lbrace \int_{-\infty}^tdt_1
\int{d\epsilon\over 2\pi}
\nonumber\\
&& \times e^{-i\epsilon(t_1-t)}{\mathbf \Gamma}^R(\epsilon,t_1,t){\mathbf
G}^r(t,t_1)
\bigl [f_L(\epsilon)-f_R(\epsilon)\bigr ]
\Bigr\rbrace \biggr ]\;.
\end{eqnarray}
The time-average of a time-dependent object $F(t)$
is defined by
\begin{equation}\label{avedef}
\big \langle F(t) \big \rangle
= \lim_{T\to\infty} {1\over T} \int_{-T/2}^{T/2} dt F(t) \;.
\end{equation}
If $F(t)$ is a periodic function of time, it is sufficient
to average over the period.
Upon time-averaging, the first term in Eq.(\ref{eqn}) vanishes,
$\big\langle dN /dt \big\rangle
\rightarrow 0$,
because the occupation $N(t)$ is finite for all $T$.
The expression for the
time-averaged current further simplifies
if one can factorize the energy- and time-dependence
of the tunneling coupling,
$V_{k\alpha,n}(t) \equiv u(t)V_{\alpha,n}(\epsilon_k)$.
We then obtain

\begin{eqnarray}\label{jave}
\big\langle J_L(t) \big\rangle = &&
 - {2e\over\hbar} \int {d\epsilon\over
2\pi}[f_L(\epsilon)-f_R(\epsilon)]\nonumber\\
&& \times {\mathrm{ ImTr}}\Bigl\lbrace
{{\mathbf{\Gamma}}^L(\epsilon){\mathbf{\Gamma}}^R(\epsilon) \over
{\mathbf{\Gamma}}^L(\epsilon)+{\mathbf{\Gamma}}^R(\epsilon)}
\big\langle u(t){\mathbf A}(\epsilon,t)\big\rangle
\Bigr\rbrace\;,
\end{eqnarray}
where
\begin{eqnarray}\label{bfA}
{\mathbf{A}}(\epsilon,t) = && \int dt_1 u(t_1){\mathbf{G}}^r(t,t_1)
\nonumber\\
&& \times\exp [ i\epsilon(t-t_1)+i\int_{t_1}^tdt_2\Delta(t_2) ]
\;.
\end{eqnarray}
Due to Eq.(\ref{propcond}) we do not have to distinguish between $L/R$ in the
definition of ${\mathbf{A}}(\epsilon,t)$; below we shall encounter situations
where this distinction is necessary.

The expression (\ref{jave}) is of the Landauer type:
it expresses the current as an integral over
a weighted density of states times the difference
of the two contact occupation factors.  It is valid
for arbitrary interactions in the central region, but
it was derived with the somewhat restrictive
assumption of proportional couplings to the leads.

\section{Non-interacting resonant-level model}

\subsection{General formulation}

In the non-interacting case the Hamiltonian
is $H=H_c+H_T+H_{\mathrm{ cen}}$, where
$H_{\mathrm{ cen}}=\sum_n\epsilon_n
{\mathbf{d}}_n^{\dagger}{\mathbf{d}}_n$.  Following
standard analysis (an analogous calculation is also carried
out in  Appendix B), one can derive the Dyson equation for
the retarded Green function,
\begin{eqnarray}\label{Gret}
{\mathbf G}^r && (t,t') \,  = {\mathbf g}^r(t,t')
\nonumber\\&& +
\int dt_1 \int dt_2 {\mathbf g}^r(t,t_1)
{\mathbf \Sigma}^r(t_1,t_2){\mathbf G}^r(t_2,t')\;,
\end{eqnarray}
where
\begin{equation}
\Sigma^r_{nn'}(t_1,t_2)  = \sum_{k\alpha\in L,R}
V^*_{k\alpha,n}(t_1)g^r_{k\alpha}(t_1,t_2)V_{k\alpha,n'}(t_2)\;,
\end{equation}
and $g^r_{k\alpha}$ is given by Eq.(\ref{glead}).
{}From (\ref{keldysh}) the Keldysh equation for $G^<$ is\cite{firstterm}
\begin{eqnarray}\label{Gless}
{\mathbf G}^<(t,t') \, && =\int dt_1 \int dt_2 {\mathbf G}^r(t,t_1)
{\mathbf \Sigma}^<(t_1,t_2){\mathbf G}^a(t_2,t') \nonumber \\
&& = i\int dt_1\int dt_2 {\mathbf G}^r(t,t_1)
\biggl [ \sum_{L,R} \int {d\epsilon\over 2\pi}
{\mathrm{ e}}^{-i\epsilon(t_1-t_2)}\nonumber\\
&& \times f_{L/R}(\epsilon){\mathbf \Gamma}^{L/R}(\epsilon,t_1,t_2) \biggr ]
{\mathbf G}^a(t_2,t') \;.
\end{eqnarray}
Provided that the Dyson equation for the retarded Green function can be solved,
Eq.(\ref{Gless}) together with the current expression Eq.(\ref{curfin})
provides
the complete solution to the noninteracting resonant-level model.  Below we
examine special cases where analytic progress can be made.

\subsection{Time-independent case}

In the time-independent case the time-translational invariance is restored,
and it is advantageous to go over to energy variables:
\begin{eqnarray}
{\mathbf G}^r(\epsilon) \, && \, =[({\mathbf
g}^r)^{-1}-{\mathbf\Sigma}^r(\epsilon)]^{-1}
\nonumber \\
{\mathbf G}^<(\epsilon) \, && \, ={\mathbf
G}^r(\epsilon){\mathbf\Sigma}^<(\epsilon)
{\mathbf G}^a( \epsilon) \;.
\end{eqnarray}
In general the Dyson equation for the retarded Green function
requires matrix inversion.  In the case of a single level, the scalar equations
can be readily solved.
The retarded (advanced) self-energy is
\begin{equation}\label{sigma}
\Sigma^{r,a}(\epsilon) = \sum_{k\alpha\in L,R}
{|V_{k\alpha}|^2\over\epsilon-\epsilon_{k\alpha}\pm i\eta}
=\Lambda(\epsilon) \mp {i\over 2}\Gamma(\epsilon)\;,
\end{equation}
where the real and imaginary parts contain 'left' and 'right' contributions:
$\Lambda(\epsilon)=\Lambda^L(\epsilon)+\Lambda^R(\epsilon)$ and
$\Gamma(\epsilon)=\Gamma^L(\epsilon)+\Gamma^R(\epsilon)$.
The lesser self-energy is
\begin{eqnarray}
\Sigma^<(\epsilon) = && \sum_{k\alpha\in L,R} |V_{k\alpha}|^2
g_{k\alpha}^<(\epsilon)\nonumber\\
= && i[\Gamma^L(\epsilon)f_L(\epsilon)+\Gamma^R(\epsilon)f_R(\epsilon)]\;.
\end{eqnarray}
Using  the identities
$G^rG^a=(G^r-G^a)/(1/G^a-1/G^r)=a(\epsilon)/\Gamma(\epsilon)$
[here $a(\epsilon)=i[G^r(\epsilon)-G^a(\epsilon)]$ is the spectral function],
one
can write $G^<$ in a 'pseudoequilibrium' form:
\begin{equation}
G^<(\epsilon) = i a(\epsilon) {\bar f}(\epsilon)\;,
\end{equation}
where
\begin{eqnarray}
{\bar f}(\epsilon) && =
{\Gamma^L(\epsilon)f_L(\epsilon)+\Gamma^R(\epsilon)f_R(\epsilon)
\over \Gamma(\epsilon) }\nonumber\\
\nonumber\\
a(\epsilon) && = {\Gamma(\epsilon)\over
[\epsilon-\epsilon_0- \Lambda(\epsilon)]^2 +[\Gamma(\epsilon)/2]^2}\;.
\end{eqnarray}
With these expressions the evaluation of the current (\ref{Ned92})
is straightforward:
\begin{eqnarray}\label{jdc}
J  = -{e\over 2\hbar}\int {d\epsilon\over 2\pi} && a(\epsilon)
\Bigl [ \bigl ( \Gamma^L(\epsilon)-\Gamma^R(\epsilon) \bigr ) {\bar
f}(\epsilon)\nonumber\\
&& -\bigl ( f_L(\epsilon)\Gamma^L(\epsilon)-f_R(\epsilon)\Gamma^R)\bigr ) \Bigr
]
\nonumber\\
\nonumber\\
 = {e\over\hbar}\int{d\epsilon\over 2\pi} &&
{\Gamma^L(\epsilon)\Gamma^R(\epsilon)\over
[\epsilon-\epsilon_0- \Lambda(\epsilon)]^2 +[\Gamma(\epsilon)/2]^2}
\nonumber\\
\nonumber\\
&& \times [f_L(\epsilon) - f_R(\epsilon)]\;.
\end{eqnarray}
Note that this derivation made no assumptions about proportionate
coupling to the leads.
The factor multiplying the difference of the fermi functions is the
elastic transmission coefficient.
It is important to understand the difference
between this result and the result obtained
in Section III.B.2 (despite the similarity
of appearance):  There Eq.(\ref{jdcprop}) gives the current for
a fully interacting system, and the evaluation of the retarded
and advanced Green functions requires a consideration of
interactions (e.g., electron-electron, electron-phonon, and
spin-flip) in addition to tunneling back
and forth to the contacts.
{\it Suppose} now that the Green function for the interacting central region
can be solved:
$G^{r,a}(\epsilon) = [\epsilon - \epsilon_0 - \lambda(\epsilon)
\pm i \gamma(\epsilon)/2]^{-1}$, where $\lambda$ and $\gamma/2$ are the
real and imaginary parts of the self-energy (including interactions
{\it and} tunneling).  Then the interacting result for proportionate
coupling (\ref{jdcprop}) becomes
\begin{eqnarray}
J = {e\over\hbar}\int{d\epsilon\over 2\pi} && [f_L(\epsilon) - f_R(\epsilon)]
{\Gamma^L(\epsilon)\Gamma^R(\epsilon)\over
\Gamma^L(\epsilon)+\Gamma^R(\epsilon)}
\nonumber\\
\nonumber\\
&& \times {\gamma(\epsilon)\over
[\epsilon-\epsilon_0- \lambda(\epsilon)]^2 +[\gamma(\epsilon)/2]^2}\;.
\end{eqnarray}
This result coincides with the noninteracting current expression
(\ref{jdc}) if $\lambda(\epsilon)\to\Lambda(\epsilon)$ and
$\gamma(\epsilon)\to\Gamma(\epsilon)=\Gamma^R(\epsilon)+\Gamma^L(\epsilon)$.
In a phenomenological model, where the total level width is expressed
as a sum of elastic and inelastic widths, $\gamma=\gamma_e+\gamma_i$,
one recovers the results of Jonson and Grincwajg, and Weil and Vinter.
\cite{JGWV}

\subsection{Wide-band limit}\label{WBLsec}
\subsubsection{Basic formulae}
For simplicity, we continue to consider only a single level in the central
region.
As in the previous section, we assume
that one can factorize the momentum and time-dependence of the tunneling
coupling, but allow for
different time-dependence for each barrier:
$V_{k\alpha}(t) \equiv u_{L/R}(t)V_{\alpha,n}(\epsilon_k)$.
Referring to Eq.(\ref{sigma}),
the wide-band limit (WBL) consists of i) neglecting the level-shift
$\Lambda(\epsilon)$, ii) assuming that the line-widths are
energy independent constants, $\sum_{\alpha\in L,R}\Gamma_{\alpha} =
\Gamma^{L/R}$,
and (iii) allowing a single
time-dependence, $\Delta_{L/R}(t)$, for the energies
in each lead.
The retarded self-energy in Eq.(\ref{Gret})
thus becomes
\begin{eqnarray}\label{sigmarestun}
\Sigma^r(t_1,t_2) \, && =
\sum_{\alpha\in L,R} u^*_{\alpha}(t_1)u_{\alpha}(t_2)
{\mathrm{ e}}^{-i\int_{t_2}^{t_1}dt_3\Delta_{\alpha}(t_3)}\nonumber\\
&& \times\int{d\epsilon\over 2\pi}{\mathrm{ e}}^{-i\epsilon(t_1-t_2)}
\theta(t_1-t_2)[-i\Gamma_{\alpha}]
\nonumber \\
&& = -{i\over 2}[\Gamma^L(t_1)+\Gamma^R(t_1)]\delta(t_1-t_2)\;.
\end{eqnarray}
(Here we have introduced the notation
$\Gamma^{L/R}(t_1)\equiv\Gamma^{L/R}(t_1,t_1)=\Gamma^{L/R} |u_{L/R}(t_1)|^2$.)
With this self-energy, the retarded(advanced) Green function becomes
\cite{LangNord,ned}
\begin{equation}\label{Greslev}
G^{r,a}(t,t') = g^{r,a}(t,t')
\exp{\Bigl\lbrace\mp\int_{t'}^{t}dt_1{1\over 2}
\bigl [ \Gamma^L(t_1)+\Gamma^R(t_1)\bigr ]\Bigr\rbrace}
\end{equation}
with
\begin{equation}
g^{r,a}(t,t')=\mp i\theta(\pm t\mp t')
\exp{\Bigl[-i\int_{t'}^{t}dt_1\epsilon_0(t_1)\Bigr]}\;.
\end{equation}
This solution can now be used to evaluate the
lesser function Eq.(\ref{Gless}), and further in
Eq.(\ref{curfin}), to obtain the time-dependent current.  In the WBL the
$\epsilon$-
and $t_1$-integrals
in the term involving $G^<$ are readily evaluated, and we write the
current as (using ${\mathrm{ Im}} \lbrace
G^<(t,t) \rbrace = N(t)$, where $N(t)$ is the occupation of the
resonant level)
\begin{eqnarray}
J_L(t) && =-{e\over\hbar}\Bigl [ \Gamma^L(t) N(t)
+
\int {d\epsilon\over\pi} f_L(\epsilon)\nonumber\\
&& \times\int_{-\infty}^t dt_1 \Gamma^L(t_1,t)
{\mathrm{ Im}} \lbrace
{\mathrm{ e}}^{-i\epsilon(t_1-t)}
G^r(t,t_1)\rbrace \Bigr ]\;.
\end{eqnarray}
For a compact notation we introduce
\begin{eqnarray}\label{A}
A_{L/R}(\epsilon,t) = && \int dt_1 u_{L/R}(t_1) G^r(t,t_1)\nonumber\\
&& \times\exp [ i\epsilon(t-t_1)-i\int_t^{t_1}dt_2\Delta_{L/R}(t_2) ]
\;.
\end{eqnarray}
Obviously, in the time-independent case  $A(\epsilon)$ is just the Fourier
transform of
the retarded Green function $G^r(\epsilon)$.\cite{Aversusa}
In terms of $A(\epsilon,t)$ the occupation $N(t)$ (using Eq.(\ref{Gless})
for $G^<$)
is given by
\begin{equation}\label{occup}
N(t) = \sum_{L,R}\Gamma^{L/R}
\int {d\epsilon \over 2 \pi} f_{L/R}(\epsilon)
|A_{L/R}(\epsilon,t)|^2\;.
\end{equation}
We write the current as a
sum of currents flowing out from the central region into the
left(right) contact (see also Fig.\ \ref{fig9}), and  currents flowing
into the central
region from the left(right) contact,
$J_{L/R}(t)=J_{L/R}^{\mathrm{out}}(t) +
J_{L/R}^{\mathrm{in}}(t)$:\cite{sign}
\begin{eqnarray}\label{jonelevel}
J_{L/R}^{\mathrm{out}}(t) &&
= -{e\over\hbar}  \Gamma^{L/R}(t)N(t)\\
J_{L/R}^{\mathrm{in}}(t) && =
-{e\over\hbar}\Gamma^{L/R}u_{L/R}(t) \int{d\epsilon\over\pi}
f_{L/R}(\epsilon) {\mathrm{ Im}}\lbrace A_{L/R}(\epsilon,t)\rbrace \;.
\end{eqnarray}
It is readily verified that these expressions coincide with
Eq.(\ref{jdc}) if all time-dependent quantities are replaced
by constants.

Employing the same approach as in Section IV C,
and provided that $u_L(t)=u_R(t)=u(T)$,
we find that the
time-averaged current in the WBL is given by
\begin{eqnarray}\label{WBLAVE}
\langle J \rangle =
- {2e \over \hbar} { {\Gamma^L \Gamma^R} \over {\Gamma^L + \Gamma^R} }
\int { {d\epsilon} \over {2 \pi} } && {\mathrm{Im}} \{
   f_L(\epsilon) \langle u(t) A_L(\epsilon,t) \rangle \nonumber\\
&& - f_R(\epsilon) \langle u(t) A_R(\epsilon,t) \rangle \}\;.
\end{eqnarray}
Unlike the general case of Eq. (\ref{jave}), there is no
restriction in the WBL that the energy dependence be the same in
the two leads. Eq. (\ref{WBLAVE}) can therefore be used for the
case of a time-dependent bias, where $\Delta_L(t)$ and $\Delta_R(t)$
will be different. It is interesting to note that the function
of energy appearing in the time-averaged current is positive
definite. In particular, as is shown in Appendix C,
\begin{equation}\label{NEGIMA}
-\langle{\mathrm{Im}} \{ u_{L/R}(t)A_{L/R}(\epsilon,t) \}\rangle
= {\Gamma\over 2}\langle|A_{L/R}(\epsilon,t)|^2\rangle\;.
\end{equation}
One consequence of (\ref{NEGIMA})
is that if only the level is time dependent
the average current cannot flow against the bias.

In the next two sections we consider specific
examples for the time variation, which are relevant for
experimental situations.

\subsubsection{Response to harmonic modulation}
Harmonic time modulation is probably the most commonly
encountered example of time dependence. Here we treat the case
when the contact and site energy levels vary as
\begin{equation}\label{delta}
\Delta_{L/R,0}(t)=\Delta_{L/R,0}\cos(\omega t)
\end{equation}
It is easy to generalize the treatment to situations where
the modulation frequencies and/or phases are different
in different parts of the device.
Assuming that the barrier heights do
not depend on time ($u_{L/R}=1$), and
substituting (\ref{delta})
in the expression (\ref{A}) for $A(\epsilon,t)$, one
finds \cite{bessel}
\begin{eqnarray}\label{Aharmonic}
A_{L/R} && (\epsilon,t) = \exp
{\bigl [-i{(\Delta_0-\Delta_{L/R})\over\omega}\sin(\omega t)\bigr ]}
\nonumber\\
&&\times\sum_{k=-\infty}^{\infty}
J_k\biggl({\Delta_0-\Delta_{L/R}\over\omega}\biggr)
{e^{ik\omega t} \over \epsilon - \epsilon_0-k\omega+i\Gamma/2}\;,
\end{eqnarray}
where $J_{-k}(x) = (-1)^k J_k(x)$.
Figures\ \ref{fig3} (a) and (b) show $|A(\epsilon,t)|^2$ and
${\mathrm{Im}}A(\epsilon,t)$ as a function of time, respectively.
We recall from Eqs.(\ref{occup}-\ref{jonelevel}) that the
current at a given time is determined by integrating
$|A(\epsilon,t)|^2$ and
${\mathrm{Im}}A(\epsilon,t)$ over energy, and thus an examination
of Figure\ \ref{fig3} helps to understand to complicated time dependence
discussed below.
(We show only $A_L$; similar results hold for $A_R$.)
The physical parameters used to generate these plots are given in the figure
caption.
The three-dimensional plot (top part of figure) is projected
down on a plane to yield a contour plot in order to help
to visualize the time dependence.  As expected, the time
variation is periodic with period $T=2\pi/\omega$.  The time dependence
is strikingly complex.  The most easily recognized features are
the maxima in the plot for $|A|^2$; these are related to
photon side-bands occuring at $\epsilon
=\epsilon_0\pm k\omega $ (see also Eq.(\ref{aharmave}) below).\cite{BL}

The current is computed using the methods
described in Appendix B, and is shown in Fig.\ \ref{fig4}.
We also display the drive voltage as a broken line.  Bearing
in mind the complex time dependence of $|A|^2$ and ${\mathrm{Im}}A$,
which determine the out- and in-currents, respectively, it is
not surprising that the current displays a non-adiabatic time dependence.
The basic physical mechanism underlying the secondary maxima and
minima in the current is the line-up of a photon-assisted
resonant-tunneling peak with the contact chemical potentials.
The rapid time variations are due to $J^{\mathrm in}$ (or, equivalently,
due to ${\mathrm{Im}}A$): the out-current  $J^{\mathrm out}$ is determined
by the occupation $N(t)$, and hence varies only on a time-scale $\Gamma/\hbar$,
which is the time scale for charge density changes.

We next consider the time-average of the current.
For the case of harmonic time dependence,
we find\cite{bessel}
\begin{eqnarray}\label{aharmave}
\big\langle {\mathrm{Im}} A_{L/R}  (\epsilon,t)\big\rangle  && =
-{\Gamma\over 2}
\sum_{k=-\infty}^{\infty}
J_k^2\biggl({\Delta_0-\Delta_{L/R}\over\omega}\biggr)
\nonumber\\
&& \times {1 \over (\epsilon - \epsilon_0-k\omega)^2+(\Gamma/2)^2}\;.
\end{eqnarray}
Figure \ref{fig5} shows the resulting time-averaged current $J_{\mathrm{dc}}$.
A consequence of the complex harmonic structure of the time-dependent
current is that for temperatures $k_BT<\hbar\omega$ the average current
oscillates as function of period $2\pi/\omega$.  The oscillation
can be understood by examining the general expression for average
current Eq.(\ref{jave}) together with (\ref{aharmave}):  whenever a
photon-assisted peak in the effective density of states, occuring
at $\epsilon=\epsilon_0\pm k\omega $
in the time-averaged density of
states $\langle{\mathrm{Im}}A_{L/R}$, moves in or out of the
allowed energy range, determined by the difference of the
contact occupation factors, a maximum (or minimum) in the
average current results.

\subsubsection{Response to steplike modulation}\label{WBLstep}
We give results for the case when the central site level
changes abruptly at $t=t_0$: $\epsilon_0 \to \epsilon_0 + \Delta$.
If the contacts also change at the same time, the corresponding
results are obtained by letting $\Delta \to \Delta - \Delta_{L/R}$.
Thus, simultaneous and equal shifts
in the central region and the contacts
have no effect.  Assuming that the barrier heights
do not depend on time ($u_{L/R}\equiv 1$),
one finds for $t>t_0$ from Eq.(\ref{A})
\begin{eqnarray}\label{Aabrupt}
A && (\epsilon,t) =  { 1 \over \epsilon - \epsilon_0 +i\Gamma/2}
\nonumber\\
&& \times\Bigl\lbrace 1 + \Delta{
[1-\exp{[i(\epsilon-(\epsilon_0+\Delta)+i\Gamma/2)(t-t_0)}]
\over \epsilon-(\epsilon_0+\Delta)+i\Gamma/2}
\Bigr\rbrace\;.
\end{eqnarray}
This result is easily generalized (See Eq.(14) in Ref.[\onlinecite{WJM93}]) to
a pulse
of duration $s$, and
numerical results are discussed below.

It is instructive to study analytically
the long and short-time behavior of $A(\epsilon,t)$.
It easily verified that $A(\epsilon,t)$ has the expected limiting
behavior:
\begin{equation}
A(\epsilon,t \to \infty) = [\epsilon-(\epsilon_0+\Delta)+i\Gamma/2]^{-1}\;.
\end{equation}
Thus, when the transients have died away, $A(\epsilon,t)$ settles
to its new steady-state value.

Consider next the change in current at short times after the pulse,
$t-t_0 \equiv \delta t \ll \hbar/\Gamma,
\hbar/\epsilon$.
Note that the second inequality provides an effective cut-off
for the energy integration required for the current.
In this limit we may write
\begin{equation}
A(\epsilon,t)\simeq {1-i\Delta\delta t \over
\epsilon-\epsilon_0+i\Gamma/2}.
\end{equation}
Since $\delta J^{\mathrm{ out}}(t) \propto |A(\epsilon,t)|^2 \propto (\delta
t)^2$,
the leading contribution comes from $J^{\mathrm{ in}}(t)$.
For low temperatures we find
\begin{eqnarray}
\delta J_{L/R}(t) \simeq && {e\Gamma^{L/R}\over \pi\hbar}
\int_{-\hbar/\delta t}^{\mu_{L/R}} d\epsilon
{\mathrm{ Im}}\delta A(\epsilon,t)\nonumber\\
&& \simeq
{e\Gamma^{L/R}\over \pi\hbar}
\Delta\delta t \log \delta t\;.
\end{eqnarray}

We next discuss the numerical results for a step-like modulation.
Just like in the case of harmonic modulation, it is
instructive to study the time dependence of $|A|^2$
and ${\mathrm{Im}}A$; these are shown in Figures \ref{fig6}(a)
and (b), respectively.  The observed time dependence
is less complex than in the harmonic case.
Nevertheless, the resulting current,
which we have computed for a pulse of duration $s$, and
display in Fig.\ \ref{fig7},
shows an interesting ringing behavior.   The ringing is
again due to the movement of the sidebands of
${\mathrm{Im}}A_{L/R}$ through the contact Fermi energies.

Due to the experimental caveats discussed in Section II, the
ringing showed in Fig.\ \ref{fig7} may be masked by
capacitive effects not included in the present work.  However,
the ringing should be observable in the time-averaged
current by applying a series of pulses such as that of
Fig.\ \ref{fig7}, and then varying the pulse duration.\cite{Leo}
In Fig.\ \ref{fig8} the derivative of the dc current with
respect to pulse length is plotted, normalized by
the repeat time $\tau$ between pulses.  For pulse
lengths $s$ of the order of the resonance liftime
$\hbar/\Gamma$, the derivative of the dc current
mimics closely the time-dependent current following
the pulse, and, likewise, asymptotes to the
steady-state current at the new voltage.

\subsubsection{Linear-response}

For circuit modeling purposes it would often
be desirable to replace the mesoscopic device
with a conventional circuit element, with an
associated complex impedance $Z(\omega)$, or
admittance $Y(\omega)$. Our results for the nonlinear
time-dependent current form a very practical starting point for
such a calculation.   For the non-interacting
case, the current is determined by $A(\epsilon,t)$
(see Eq.(\ref{occup}-\ref{jonelevel})), and all one has to do
is to linearize $A$ (Eq.(\ref{A})) with respect
to the amplitude of the drive signal, {\it i.e.},
$\Delta-\Delta_{L/R}$.  It is important to notice
that we do not linearize with respect to the
chemical potential difference: the results given
below apply to an arbitrary static bias voltage.

Performing the linearization, one finds
\begin{eqnarray}\label{A2lin}
|A_{L/R}^{(1)} && (\epsilon,t)|^2 = {\Delta-\Delta_{L/R}\over\omega}
{\mathrm{Re}}\Bigl\lbrace
{1\over \epsilon-\epsilon_0+i\Gamma/2}\nonumber\\
&&\times\bigl[{{\mathrm{e}}^{-i\omega
t}\over\epsilon-\epsilon_0-\omega-i\Gamma/2}
-{{\mathrm{e}}^{i\omega t}\over\epsilon-\epsilon_0+\omega-i\Gamma/2}\bigr]
\Bigr\rbrace\;,
\end{eqnarray}
and
\begin{eqnarray}\label{ImAlin}
{\mathrm{Im}}A_{L/R}^{(1)} && (\epsilon,t) = {\Delta-\Delta_{L/R}\over 2\omega}
{\mathrm{Im}}\Bigl\lbrace
{{\mathrm{e}}^{i\omega t}\over\epsilon-\epsilon_0-\omega+i\Gamma/2}\nonumber\\
&& -{{\mathrm{e}}^{-i\omega t}\over\epsilon-\epsilon_0+\omega+i\Gamma/2}
+{{\mathrm{e}}^{-i\omega t}-{\mathrm{e}}^{i\omega t}
\over \epsilon-\epsilon_0+i\Gamma/2}
\Bigr\rbrace\;.
\end{eqnarray}
At finite temperature the energy integration must be done numerically,
as explained in Appendix B, while at $T=0$ they can be done analytically.
In the latter case, all the integrals can be cast into the form
\begin{eqnarray}\label{integrals}
\int_{-\infty}^{\mu} &&
{d\epsilon \over (\epsilon-\epsilon_1+i\Gamma_1/2)
(\epsilon-\epsilon_2+i\Gamma_2/2)}
= \nonumber\\
&& {1\over \epsilon_1-\epsilon_2+i(\Gamma_2-\Gamma_1)/2}
\log {\mu-\epsilon_1+i\Gamma_1/2 \over
\mu-\epsilon_2+i\Gamma_2/2 }\;.
\end{eqnarray}
Using $\log(x+iy)=1/2\log(x^2+y^2)+i\tan^{-1}(y/x)$ yields
\begin{eqnarray}\label{lincurin}
J^{(1),{\mathrm{in}}}_{L/R} =   {e\over\hbar}\Gamma^{L/R} &&
{\Delta-\Delta_{L/R}\over 2\pi\omega}
\bigl[\cos(\omega t) F_{L/R}(\omega)\nonumber\\
&& + \sin(\omega t) G_{L/R}(\omega)\bigr]
\end{eqnarray}
and
\begin{eqnarray}\label{lincurout}
&& J^{(1),{\mathrm{out}}}_{L/R} =  {e\over\hbar}\Gamma^{L/R}
\sum_{L,R}\Gamma^{L/R}{\Delta-\Delta_{L/R}\over 2\pi\omega}\nonumber\\
&&\Bigl\lbrace \cos(\omega t)\bigl[{\omega\over\omega^2+\Gamma^2}
G_{L/R}(\omega)
-{\Gamma\over\omega^2+\Gamma^2}F_{L/R}(\omega)\bigr]\nonumber\\
&& -\sin(\omega t)\bigl[{\Gamma\over\omega^2+\Gamma^2}G_{L/R}(\omega)
+{\omega\over\omega^2+\Gamma^2}F_{L/R}(\omega)\bigr]\Bigr\rbrace ,\nonumber\\
\end{eqnarray}
where we defined
\begin{equation}\label{Gomega}
G_{L/R}(\omega) =  \log {|\mu_{L/R}-\epsilon_0 + i\Gamma/2|^2
\over
|(\mu_{L/R}-\epsilon_0+i\Gamma/2)^2-\omega^2|}
\end{equation}
and
\begin{eqnarray}\label{Fomega}
F_{L/R}(\omega) = && \tan^{-1} {\mu_{L/R}
-\epsilon_0-\omega\over\Gamma/2} \nonumber\\
&& - \tan^{-1} {\mu_{L/R}-\epsilon_0+\omega\over\Gamma/2} \;.
\end{eqnarray}
These expressions give the linear ac current for an arbitrarily
biased double barrier structure, where both contacts and the central
region energies are allowed to vary harmonically.
As a check, it is instructive to verify that the finite temperature
results of Appendix B.2 contain Eqs.(\ref{lincurin}-\ref{lincurout}) as a
special case;
this is a rather straightforward calculation using the limiting behavior
of the Digamma function.

Considerable simplification
occurs, if one considers a symmetric structure at zero bias:
$\Gamma^L = \Gamma^R = \Gamma/2$, and
$\mu_L = \mu_R \equiv \mu$, respectively.
Following Fig.\ \ref{fig9}, the net current from left to right is
$J^{(1)}= 1/2[J^{(1),{\mathrm{in}}}_L+J^{(1),{\mathrm{out}}}_R-
J^{(1),{\mathrm{out}}}_L-J^{(1),{\mathrm{in}}}_R]$.  Using
Eq.(\ref{lincurin}-\ref{lincurout}),
one finds that the 'out' contributions cancel, and that the
'in'-currents combine to give the
net current
\begin{equation}\label{lincursym}
J^{(1)}(t) = -{e\over\hbar}{\Gamma\over 4}
{\Delta_L-\Delta_R\over 2\pi\omega}
[\cos(\omega t)F(\omega)+\sin(\omega t)G(\omega)]\;.
\end{equation}
Here the functions $F(\omega)$ and $G(\omega)$ are given by
Eq.(\ref{Fomega}-\ref{Gomega}), but using $\mu$ and $\Gamma/2$
as parameters.  This result exactly coincides with the recent
calculation of Fu and Dudley,\cite{fududley} which employed
a different method.

We now wish to apply the formal results derived in
this section to an experimentally relevant system.
The archetypal mesoscopic with potential for applications
is the resonant-tunneling diode.
The key feature of a resonant-tunneling diode is
its ability to show negative differential resistance (NDR).
The WBL model studied in this section does not have this
feature: its IV-characteristic, which is readily
evaluated with Eq.(\ref{jdc}), is a monotonically
increasing function.  A much more interesting model
can be constructed by
considering a model where the
contacts have a finite occupied bandwidth; this can
achieved by introducing
a low energy cut-off $D_{L/R}$ (in addition to the upper
cut-off provided by the electro-chemical potential).
The zero-temperature IV-characteristic is now
\begin{eqnarray}\label{ivfake}
&& J_{\mathrm{dc}}(V) = {e\over h}{2\Gamma_L \Gamma_R\over \Gamma}
\bigl[
\tan^{-1}{\mu_L-\epsilon_0(V)\over\Gamma/2}
-\tan^{-1}{\mu_L-D_L-\epsilon_0(V)\over\Gamma/2}\nonumber\\
&& -\tan^{-1}{\mu_R(V)-\epsilon_0(V)\over\Gamma/2}
+\tan^{-1}{\mu_R(V)-D_R-\epsilon_0(V)\over\Gamma/2}\bigr]\;.
\nonumber\\
\end{eqnarray}
Here we assume that the right chemical potential is field
dependent: $\mu_R(V)=\mu_R-eV$, and that
the field-dependence of the
central region level is given by $\epsilon_0(V)=\epsilon_0
-V/2$.
The resulting current-voltage characteristic is depicted in
Fig.\ \ref{fig10}.  We note that the strong increase
in current,  which is observed in experimental systems
at very high voltages, is not present in our
model: this is because we have ignored the bias-dependence
of the barrier heights as well as any higher lying resonances.
The only generalization required for Eqns.(\ref{lincurin}
-\ref{lincurout}) is to modify the $F$ and $G$ functions:
$F_{\mu}\to{\tilde F}=F_{\mu}-F_{\mu-D}$, and analogously for $G_{\mu}$.
We show in Fig.\ \ref{fig11} the resulting linear-response
admittance $Y(\omega)$ for a symmetric structure ($\Gamma_L
=\Gamma_R$).
Several points are worth noticing.
For dc bias $eV = 5$ (energies are given in units of $\Gamma$)
the calculated admittance
resembles qualitatively  the results reported by
Fu and Dudley for zero external bias, except that the change in sign for
the imaginary part of $Y(\omega)$
is not seen.
For zero external bias
(not shown in the figure) our finite band-width
model leads to an admittance, whose imaginary part
changes sign, and thus the behavior found by Fu and
Dudley cannot be ascribed to an artefact of their
infinite band-width model.
More interestingly, for dc bias in the NDR regime, the
real part is negative for small frequencies.  This simply
reflects the fact that the device is operating
under NDR bias conditions.  At higher frequencies the
real part becomes positive, thus indicating that further modeling along the
lines sketched here may lead to important implications
on the high-frequency response of resonant-tunneling
structures.

In concluding this section, we wish to emphasize that the linear-response
analysis presented above is only a special case of the general
results of Section IV, which seem to have the potential for
many applications.

\section{Resonant tunneling with electron-phonon interactions}
As a final application, we establish a connection to previous
calculations on the effect of phonons on resonant tunneling.
\cite{ned}  For simplicity, we consider
a single resonant level with
energy-independent
level widths $\Gamma_L$ and $\Gamma_R$ (i.e. the WBL).
The expression for the current Eq.(\ref{jdcprop}) becomes now
\begin{equation}\label{jconstg}
J = {e\over\hbar}{\Gamma^L\Gamma^R \over \Gamma^L+\Gamma^R}
\int {d\epsilon\over 2\pi}
[f_L(\epsilon)-f_R(\epsilon)]\int_{-\infty}^{\infty}dt e^{i\epsilon t}a(t)\;,
\end{equation}
where $a(t)
=i[G^r(t)-G^a(t)]$ is the interacting
spectral density. In general, an exact evaluation of $a(t)$
is not possible, however, if one {\it ignores the Fermi sea}, $G^r(t)$
(and hence $a(t)$) can be calculated
exactly:\cite{Mahan}
\begin{equation}\label{Gphonon}
G^r(t)  = -i\theta(t) \exp [-it(\epsilon_0-\Delta)-\Phi(t)-\Gamma t/2]\;,
\end{equation}
where
\begin{equation}
\Delta  = \sum_{{\mathbf q}} {M_{{\mathbf q}}^2 \over \omega_{{\mathbf q}}}\;,
\end{equation}
and
\begin{equation}
\Phi(t)  = \sum_{{\mathbf q}}{M_{{\mathbf q}}^2
\over \omega_{{\mathbf q}}^2}[N_{{\mathbf q}}
(1-e^{i\omega_{\mathbf q}t})+(N_{\mathbf q}+1)(1-e^{-i\omega_{\mathbf q}t})]\;,
\end{equation}
and the electron-phonon interaction is given by Eq.(\ref{helph}).
When substituted in the expression for current, one recovers the result
of Ref.[\onlinecite{ned}], which originally was derived by analyzing
the much more complex two-particle Green function
$G(\tau,s,t)=\theta(s)\theta(t)
\langle d(\tau-s)d^{\dag}(\tau)d(t)d^{\dag}(0)\rangle$.
The advantage of the method presented here is that
one only needs the {\it single}-particle Green function to use
the interacting current formula (\ref{jdcprop}).
Other systematic approaches to the single-particle Green
function can therefore be directly applied to the current
({\it e.g.} perturbation theory in the tunneling Hamiltonian).

\section{Conclusions}
Here, we summarize the main results of this study.
We have derived a general formula for the time-dependent current through
an interacting mesoscopic region, Eq.(\ref{curfin}).
The current is written in terms of local Green functions.
This general expression is then examined in several special cases:
(i) It is shown how earlier results for time-independent current are
contained in it [Eqs. (\ref{Ned92}) and (\ref{jdcprop})].
(ii) An exact solution, for arbitrary time-dependence,
for a single non-interacting level coupled
to two leads is given [Eqs.(\ref{occup}-\ref{jonelevel})].
This calculation leads to a prediction of 'ringing' of current in response to
abrupt
change of bias, or in response to an ac-bias.
We believe that this prediction should be experimentally verifiable.
(iii) We derive a Landauer-like formula for the average current,
Eq.(\ref{jave}), and
apply it to a simple model of dynamical disorder.  Finally, as
applications, we discuss (iv) a.c. linear-response at arbitrary dc-bias and
finite
temperature, and (v) find a connection to earlier results on resonant tunneling
in the presence of optical phonons.

We hope that time-dependence will provide a new window
on coherent quantum-transport, and will lead to significant
new insights in the future.

\acknowledgements
We have benefitted from discussions with
several colleagues: Pavel Lipavsky, Karsten Flensberg, Ben Hu and
Leo Kouwenhoven.
One of us (N.S.W) is grateful to the Nordita Mesoscopic Programme during
the initiation of this project.
Work at U.C.S.B. was supported by NSF Grant No.
NSF-DMR-9308011, by the NSF Science and Technology Center
for Quantized Electronic Structures, Grant No. DMR 91-20007,
and by NSF, ONR, and ARO at the Center for Free Electron Laser
Studies.
\appendix
\section{Nonequilibrium Green functions}

The most important result (see, e.g., Refs.[
\onlinecite{Langreth},\onlinecite{Rammer},\onlinecite{Jauho}]) of the formal
theory
of nonequilibrium Green functions
is that the perturbation
expansion has precisely the {\it same structure} as the $T=0$ equilibrium
expansion.
Instead of a time-ordered Green function, one works with the
contour-ordered Green function,
\begin{equation}\label{Gcont}
G(\tau,\tau') = -i\langle T_C\{\psi(\tau)\psi^{\dagger}(\tau')\}\rangle\;,
\end{equation}
where the contour $C$ is shown in Fig.\ \ref{fig2}.
The contour-ordering operator $T_C$ orders the operators
following it in  the contour sense: operators with time
labels later on the contour are moved left of operators
of earlier time labels.
Thus, once the self-energy functional, $\Sigma = \Sigma[G]$, has been
specified,
the contour-ordered Green function obeys
formally the same Dyson equation as in $T=0$ theory,
\begin{equation}\label{contDyseq}
G=G_0+G_0\,\Sigma\,G\;,
\end{equation}
with the modification that internal time-integrations run
along the (complex) path discussed in section II.A.
It follows from this structural equivalence
that
one can derive equations of motion just as in the $T=0$ case, and that
the passage to nonequilibrium takes place by replacing the
time-ordered  Green functions
by contour-ordered Green functions, and by replacing
the real-time integration by an integration along the time-contour.
In practical calculations, however, the contour ordered Green functions
are inconvenient, and it is expedient to perform an analytic
continuation to the real axis.
The first step in this procedure consists of expressing the
contour ordered Green functions in terms of $2\times 2$ matrices,
whose elements are determined
by which branches of the contour the two time labels
are located on.
The four elements of the matrix Green function are
not linearly independent, and it is useful to
perform a rotation of this matrix.
A particularly convenient
set of operational rules has been given by Langreth:\cite{Langreth}
If one has an expression $A = \int BC$ on the contour (this is the
generic type of term encountered in the perturbation expansion),
then the retarded and lesser  components are given by
\begin{eqnarray}\label{rules}
A^r(t,t') && = \int dt_1 B^r(t,t_1)C^r(t_1,t') \nonumber\\
A^<(t,t') && = \int dt_1 [B^r(t,t_1)C^<(t_1,t') \nonumber\\
&& \qquad +B^<(t,t_1)C^a(t_1,t')]
\end{eqnarray}
These results are readily generalized to products involving three
(or more) Green functions or self-energies.

The equation of motion for $G^<$ can be derived
by applying the rules (\ref{rules}) to the Dyson equation for the
contour-ordered Green function.
The Dyson equation can be written either in a differential form,
or in an integral form, as in Eq.(\ref{contDyseq}).
The former leads to the Baym-Kadanoff
transport equation, while the latter (which is employed in the
present work) yields the Keldysh equation for the lesser
function:
\begin{equation}\label{keldysh}
G^< = (1+G^r\Sigma^r)G^<_0(1+\Sigma^a G^a)+G^r\Sigma^<G^a \;,
\end{equation}
where the retarded and advanced Green functions
satisfy
\begin{equation}\label{gret}
G^{r,a} = G^{r,a}_0 + G^{r,a}_0 \Sigma ^{r,a} G^{r,a}\;.
\end{equation}
The physical modeling goes in the choice of
the self-energy functional $\Sigma$, which
contains the interactions (carrier-impurity scattering,
phonon scattering, carrier-carrier scattering etc.).
Once $\Sigma$ is given, for example in terms
of diagrams, the retarded, or 'lesser' components
of the self-energy
can be worked out according
to the rules (\ref{rules}),
and one can proceed to solve the coupled equations
(\ref{keldysh}) and (\ref{gret}).

\section{Dyson equation for $G^{\lowercase{t}}_{n,k\alpha}$}
\subsection{Equation-of-motion method}

According to Appendix A it is sufficient
to consider the $T=0$
equation of motion for the time-ordered Green
function $G^t_{n,k\alpha}$:
\begin{eqnarray}\label{eqmo}
-i{\partial\over\partial t'} G^t_{n,k\alpha}(t-t')
&& = \epsilon_k  G^t_{n,k\alpha}(t-t') \nonumber\\
&& + \sum_m V^*_{k\alpha,m} G^t_{nm}(t-t')\;,
\end{eqnarray}
where we defined the central region time-ordered Green function
function $G^t_{nm}(t-t') = -i\langle T\{ {{\mathbf d}}^{\dagger}_m(t')
{{\mathbf d}}_n(t) \} \rangle $.
Note that it is crucial that the
leads be non-interacting:  had we allowed interactions in
the leads the equation of motion technique would have
generated higher order Green functions in Eq.(\ref{eqmo}),
and we would not have a closed set of equations.

We can interpret the factors multiplying $G^t_{n,k\alpha}(t-t')$
as the inverse of the contact Green function operator, and introduce
a short-hand notation:
$G^t_{n,k\alpha} g^{-1}_{k\alpha} =
\sum_m G^t_{nm} V^*_{k\alpha,m}$.
By operating with $g^t_{k\alpha}$ from right, we arrive at
\begin{equation}
G^t_{n,k\alpha}(t-t')  =
\sum_m \int dt_1G^t_{nm}(t-t_1) V^*_{k\alpha,m} g^t_{k\alpha}(t_1-t')\;.
\end{equation}
According to the rules of the nonequilibrium theory, this equation
has in nonequilibrium precisely the same form, except that the intermediate
time integration runs on the complex contour:
\begin{equation}\label{Gtautau'}
G_{n,k\alpha}(\tau,\tau') =  \sum_m \int d\tau_1 G_{nm}(\tau,\tau_1)
V^*_{k\alpha,m}(\tau_1)g_{k\alpha}(\tau_1,\tau')\;.
\end{equation}
This is Eq.(\ref{Gnktau}) of the main text.  The analytic continuation
rules (\ref{rules}) can be applied, and the desired Dyson equation
is obtained.

\subsection{$S$-matrix expansion}
We write the Green function $G_{n,k\alpha}(t,t')$  in terms
of interaction-picture operators (denoted by a tilde)
by invoking the $S$-matrix:
\begin{equation}
G_{n,k\alpha}(\tau,\tau') = -i\langle T_C \{S{\tilde d}_n(\tau)
{\tilde c}_{k\alpha}^{\dagger}(\tau')\}\rangle \;,
\end{equation}
where
\begin{equation}\label{Smatrix}
S = T_C \{ \exp[-i\int_C d\tau_1 {\tilde H}_T(\tau_1)]\}
\end{equation}
is the contour-ordered $S$-matrix, and $H_T$ is the
tunneling Hamiltonian of Section II.B2. We expand the
exponential function in (\ref{Smatrix}); the zeroth order
term does not contribute, and we find
\begin{eqnarray}
G && _{n,k\alpha}(\tau,\tau') = -i\bigl\langle T_C\Bigl\lbrace
{\tilde d}_n(\tau){\tilde c}_{k\alpha}^{\dagger}(\tau')
\sum_{n=0}^{\infty}{(-i)^{n+1}\over (n+1)!}\nonumber\\
&&
\times \bigl [ \int_C d\tau_2
\sum_{k'\alpha',m}
[V_{k'\alpha',m}(\tau_2){\tilde c}_{k'\alpha'}^{\dagger}(\tau_2)
{\tilde d}_m(\tau_2)\nonumber\\
&& +V^*_{k'\alpha',m}(\tau_2){\tilde d}_m^{\dagger}(\tau_2){\tilde
c}_{k'\alpha'}(\tau_2)]
\bigr ]^{n+1}\Bigr\rbrace\bigr\rangle\;.
\end{eqnarray}
Since, by assumption, the leads are non-interacting,
result will only be non-zero if ${\tilde c}_{k\alpha}^{\dagger}(\tau')$
is contracted with ${\tilde c}_{k\alpha}(\tau_i)$ from
one of the $n+1$ interaction terms.  The $n+1$ possible choices
cancels a factor of $n+1$ in the factorial in the denominator, leaving
\begin{eqnarray}\label{alter}
G_{n,k\alpha}(\tau,\tau') && = \sum_m\int_C d\tau_2  (-i)\langle T_C\{
{\tilde c}_{k\alpha}(\tau_2){\tilde
c}_{k\alpha}^{\dagger}(\tau')\}\rangle\nonumber\\
&& \times V^*_{k\alpha,m}(\tau_2)(-i)\langle T_C\{ S {\tilde d}_m^{\dagger}
(\tau_2)
{\tilde d}_n(\tau)\}\rangle\;.
\end{eqnarray}
Eq.(\ref{alter}) is completely equivalent to the result (\ref{Gtautau'})
obtained
in the previous subsection.

\section{Proof of Eq.(\ref{NEGIMA})}

In this Appendix, we prove that for a single level in the WBL
(see Section V C) there is a definite relation,
\begin{equation}
-\langle u_{L/R}(t) {\mathrm{Im}}\{A_{L/R}(\epsilon, t)\} \rangle
= {\Gamma\over 2} \langle |A_{L/R}(\epsilon,t)|^2 \rangle,
\label{APPCID}
\end{equation}
between the time
averages of the quantities that, respectively,  determine the current
and the occupation.
For the case of the occupation, one can explicitly write
out $\langle |A_{L/R}(\epsilon,t)|^2 \rangle$ and then use
the identity
\begin{eqnarray}
&& G^r(t,t_1)G^a(t'_1,t) = i\theta(t-t_1)\theta(t-t'_1)
\nonumber\\
&&\times \left[ e^{-\Gamma(t-t'_1)}G^r(t'_1,t_1)
 -  e^{-\Gamma(t-t_1)}G^a(t'_1,t_1) \right]
\end{eqnarray}
to obtain
\begin{eqnarray}
&& \langle|A|^2\rangle = \lim_{T \to \infty} {i\over T \Gamma}
\int_{-T/2}^{T/2}dt_1 \int_{-T/2}^{T/2}dt'_1
  u_{L/R}(t_1)u_{L/R}(t'_1)
\nonumber\\
&& \times  (G^r(t'_1,t_1)-G^a(t'_1,t_1))
\exp[i\epsilon(t'_1-t_1)+\int_{t_1}^{t_{1'}}dt_2 \Delta(t_2)].
\nonumber\\
&&
\end{eqnarray}
Writing out
$\langle u_{L/R}(t) {\mathrm{Im}}\{A_{L/R}(\epsilon, t)\} \rangle$
explicitly then yields Eq. (\ref{APPCID}).

\section{Numerical integration}
In this Appendix we describe methods to facilitate  numerical calculations
in the wide-band limit (Section \ref{WBLsec}).
While the numerical integrations required for the occupation
and for the current can be done directly,
it is often difficult to obtain sufficient accuracy.
We have found that
it is useful to do the integrations analytically
by contour integration, and then sum the resulting residues. We have also
checked for a few selected
parameter values that the two methods give identical results.

\subsection{Steplike modulation}
We illustrate the somewhat cumbersome
but straightforward formulae by giving the
expressions for the deviation of the occupation from
its asymptotic value following a steplike
modulation of the level energy (Section \ref{WBLstep}):
$\delta N(t) = N(t) - N(t=\infty)$.
We find from Eqs.(\ref{occup}) and (\ref{Aabrupt})
\begin{eqnarray}
&& \delta N(t) =
{1\over 2\pi}\Delta^2{\mathrm{ e}}^{-\Gamma(t-t_0)}
[\Gamma^LD(\mu_L)+\Gamma^RD(\mu_R)] \nonumber\\
&& - {1\over 2\pi}\Delta{\mathrm{ e}}^{-\Gamma(t-t_0)/2}
[\Gamma^L 2{\mathrm{Re}}\{E(\mu_L)\}
+\Gamma^R 2{\mathrm{Re}}\{E(\mu_R)\}]\;,
\end{eqnarray}
where
\begin{eqnarray}
D(\mu) = && \int d\epsilon {f(\epsilon)\over (\epsilon-\epsilon_0-\Delta)^2
+(\Gamma/2)^2}\cdot
{1\over (\epsilon-\epsilon_0)^2 + (\Gamma/2)^2}
\nonumber\\
E(\mu) = && \int d\epsilon [ {f(\epsilon)\over (\epsilon-\epsilon_0-\Delta)^2
+(\Gamma/2)^2}\cdot
{{\mathrm{ e}}^{i(\epsilon-\epsilon_0-\Delta)(t-t_0)}
\over \epsilon-\epsilon_0 - i\Gamma/2} \;, \nonumber\\
&&
\end{eqnarray}
where $f(\epsilon)$ is the Fermi function with chemical
potential $\mu$.
The poles of the integrands are at $\epsilon=\epsilon_0 \pm i\Gamma/2$,
$\epsilon=\epsilon_0+\Delta\pm i\Gamma/2$, and
$\epsilon=\mu\pm i{2\pi(n+1/2)/\beta}$, respectively.
Upon closing the contour in the upper-half-plane, one obtains three
different contributions; the terms arising from
$\epsilon=\epsilon_0+i\Gamma/2$ and $\epsilon=\epsilon_0+\Delta +i\gamma/2$
obviously lead to no problems, while the sum over $n$ converges
either as $n^{-4}$
(the term originating from $D(\mu)$), or as $n^{-3}\exp({-2\pi n(t-t_0)
/\beta})$
(the term due to $E(\mu)$),
and hence also converges rapidly.

\subsection{Harmonic modulation}
In principle, the calculation proceeds as in the previous section.
However, the sum over the residues,
which results from the contour integration, converges very slowly.
A typical term in the resulting lengthy expressions converges only
as $n^{-2}$.  Significantly improved convergence can be obtained
by making use of the relation
\begin{equation}
\sum_{n=0}^{\infty}{1\over(n+a)(n+b)}={1\over a-b}
[\Psi(a)-\Psi(b)]\;,
\end{equation}
where $\Psi$ is the digamma function.
In what follows, we give the results for {\it linear-response}.
The occupation [which also gives
the current flowing out from the central region via (\ref{occup})] is
\begin{eqnarray}\label{nharm}
N(t) && = {1\over 2\pi} \sum_{L,R} {\Delta_0-\Delta_{L/R}\over
\omega^2+\Gamma^2}{\Gamma^{L/R}\over\omega}
\Bigl[\sin(\omega t)\bigl(2\Gamma r_0^{L/R} \nonumber\\
&& + \omega(I_+^{L/R}
-I_-^{L/R})-\Gamma(R_+^{L/R}+R_-^{L/R})\bigr ) \nonumber \\
&& + \cos(\omega t)\bigl (-2\omega r_0^{R/L}+
\omega(R_+^{L/R}+R_-^{R/L})\nonumber\\
&&+\Gamma(I_+^{L/R}-I_-^{L/R})\bigr ) \Bigr ]\;.
\end{eqnarray}
Here
\begin{eqnarray}
I^{L/R}_{\pm} && = {\mathrm{ Im}}\Bigl[\Psi\bigl(1/2 - {\beta\over 2\pi i}
(\mu_{L/R}-\epsilon_0\mp\omega-i\Gamma/2)\bigr)\Bigr] \nonumber\\
R^{L/R}_{\pm} && = {\mathrm{ Re}}\Bigl[\Psi\bigl(1/2 - {\beta\over 2\pi i}
(\mu_{L/R}-\epsilon_0\mp\omega-i\Gamma/2)\bigr)\Bigr] \nonumber\\
r^{L/R}_0 && = {\mathrm{ Re}}\Bigl[\Psi\bigl(1/2 + {\beta\over 2\pi i}
(\mu_{L/R}-\epsilon_0+i\Gamma/2)\bigr)\Bigr] \;.
\end{eqnarray}
The current flowing into  the central region can also be expressed
in terms of similar functions:
\begin{eqnarray}
J^{\mathrm{ in}}_{L/R}(t) && = {e\over\hbar} \Gamma^{L/R}
{\Delta-\Delta_{L/R} \over 2\pi\omega}
\bigl[\cos(\omega t)(i^{L/R}_- -i^{L/R}_+)\nonumber\\
&&+\sin(\omega t)(2r^{L/R}_0-r^{L/R}_+-r^{L/R}_-)\bigr]\;,
\end{eqnarray}
with
\begin{eqnarray}
i^{L/R}_{\pm} && = {\mathrm{ Im}}\Bigl[\Psi\bigl(1/2 + {\beta\over 2\pi i}
(\mu_{L/R}-\epsilon_0\mp\omega+i\Gamma/2)\bigr)\Bigr] \nonumber\\
r^{L/R}_{\pm} && = {\mathrm{ Re}}\Bigl[\Psi\bigl(1/2 + {\beta\over 2\pi i}
(\mu_{L/R}-\epsilon_0\mp\omega+i\Gamma/2)\bigr)\Bigr]\;.
\end{eqnarray}
By recalling $\lim_{z\to\infty}\Psi(z)\to\log(z)$, it is straightforward
to check that these results reduce to the $T=0$ case discussed in the
main text.

\begin{figure}
\caption{Sketch of charge distribution in a three-dimensional
resonant-tunneling device under dc-bias $V_{\mathrm{bias}}=\mu_L-\mu_R$ with
a time-modulation of amplitude $\Delta_{L/R}$ superposed on the leads.
As argued in the text, only a tiny fraction of charge carriers participates
in setting up the voltage drop across the structure.}
\label{fig1}
\end{figure}

\begin{figure}
\caption{The complex-time contour on which nonequilibrium Green
function theory is constructed.  In the contour sense, the time
$\tau_1$ is earlier than $\tau_2$ even though its real time
projection appears larger.}
\label{fig2}
\end{figure}

\begin{figure}
\caption{(a)$|A(\epsilon,t)|^2$ as a function of time for harmonic modulation
for a symmetric structure, $\Gamma_L=\Gamma_R=\Gamma/2$.
The unit for the time-axis is $\hbar/\Gamma$, and
all energies are measured in units of $\Gamma$, with the
values $\mu_L=10$, $\mu_R=0$,
$\epsilon_0=5, \Delta=5, \Delta_L=10$, and $\Delta_R=0$.  The modulation
frequency is $\omega=2\Gamma/\hbar$.
(b) The time-dependence
of ${\mathrm{Im}}A(\epsilon,t)$ for the case shown in Fig 3(a).}
\label{fig3}
\end{figure}

\begin{figure}
\caption{The time-dependent current $J(t)$ for harmonic modulation
corresponding to the parameters of Figure 3.
The dc bias is defined via $\mu_L=10$ and $\mu_R=0$, respectively.
The dotted line
shows (not drawn to scale) the time dependence of the drive signal.
The temperature is $k_BT=0.1\Gamma$.}
\label{fig4}
\end{figure}

\begin{figure}
\caption{Time averaged current $J_{\mathrm{dc}}$ as function of
the ac oscillation period $2\pi/\omega$.  The dc amplitudes are the same as
those in Fig. 4.}
\label{fig5}
\end{figure}

\begin{figure}
\caption{
(a) $|A(\epsilon,t)|^2$ as a function of time for step-like modulation.
At $t=0$
the resonant-level energy $\epsilon_0$ suddenly decreases by $5\Gamma$.
(b) The time dependence
of ${\mathrm{Im}}A(\epsilon,t)$ for the case shown in Fig 6(a). }
\label{fig6}
\end{figure}

\begin{figure}
\caption{Time-dependent current J(t) through a symmetric
double-barrier tunneling structure in response to a
rectangular bias pulse.  Initially, the chemical
potentials $\mu_L$ and $\mu_R$ and the resonant-level energy
$\epsilon_0$ are all zero. At $t=0$, a bias pulse (dashed curve)
suddenly increases energies in the left lead by $\Delta_L=10$ and
increases the resonant-level energy by $\Delta=5$.  At $t=3$,
before the current has settled to a new steady value,
the pulse ends and the current decays back to zero.
The temperature is $k_BT=0.1\Gamma$.}
\label{fig7}
\end{figure}

\begin{figure}
\caption{Derivative of the integrated dc current
$J_{dc}$ with respect to pulse duration $s$, normalized
by the interval between pulses $\tau$.  For
pulse durations much longer than the resonance lifetime $\hbar/\Gamma$,
the derivative is just the steady-state current at the bias voltage, but
for shorter pulses the ringing response of the current is evident.}
\label{fig8}
\end{figure}

\begin{figure}
\caption{Linear-response configuration.}
\label{fig9}
\end{figure}

\begin{figure}
\caption{IV-characteristic for a model resonant-tunneling device
(quantum dot).
The system is defined by parameters
$\epsilon_0(V=0) = 2$, $\mu_L = \mu_R(V=0) = 0$, and $D_L
=D_R=2$, and the current is given in units of
$e\Gamma/h$.}
\label{fig10}
\end{figure}

\begin{figure}
\caption{In-phase and out-of-phase components of the
linear response current (in units of $e\Gamma/h$ and
normalized with the amplitude of the drive signal $\Delta_L$
to yield admittance) for two bias points, $eV=5$ (continuous
line) and $eV=10$ (dashed line).  Other parameters are
as in Figure 10.
The out-of-phase components (or, equivalently, imaginary
parts) always tend to zero for vanishing frequency, while
the in-phase component can have either a positive or negative
zero-frequency limit depending on the dc bias.}
\label{fig11}
\end{figure}

\begin{references}
\bibitem{sign} The reader may worry about the direction of the in- and
out-currents, because both terms in  (\ref{jonelevel}) apparently have the same
sign.
However, $\langle{\mathrm{Im}}A\rangle <0$ (See Appendix D for an explicit
demonstration),
and the in- and out-currents flow in opposite directions, as expected.
\bibitem{firstterm}
Note that the first term in (\ref{keldysh}) is zero, because
$[1+G^r \Sigma^r ]G^<_0 = G^r(G^r_0)^{-1}G^<_0 = 0$.
\bibitem{Aversusa}
A remark about the notation: the
reader should not confuse $A$ with a (generalized) spectral function (which
is also often denoted by $A$.  We employ the symbol
$a$ for the spectral function, i.e.
$a(\epsilon)=-2{\mathrm{Im}}\,G^r(\epsilon)$.
\bibitem{bessel}
It is useful to recall the identity
$\exp{[i\alpha\sin(\omega t)]} =
\sum_{k=-\infty}^{\infty}
{\mathrm{ e}}^{i\omega tk}\,J_k(\alpha)$,
where $J_k$ is the $k:$th order Bessel function.
\end{references}
\end{document}